\documentstyle[11pt,epsfig]{article}
\columnwidth\textwidth
\begin{document}
\begin{flushright}
BUTP-9612\\
\end{flushright}

\begin{center}
\Large{ \bf  Fermion scattering in  
domain walls with a locally dependent phase.}
\end{center}

\begin{center}
E. Torrente Lujan. \\
Inst. fur Theoretische Physik, Universitat Bern \\
Sidlerstrasse 5, 3012 Bern, Switzerland.\\
e-mail: e.torrente@cern.ch\\
\end{center}

\begin{abstract}
We consider interactions of fermions with the domain wall
bubbles produced during a first order phase transition. A 
new exact solution of the Dirac equations is obtained for 
a wall profile  incorporating a position dependent phase
factor. The reflection coefficients are obtained.

PACS: 

\end{abstract}

\newpage

\section{Introduction}

During the last few  years, a considerable amount of work has been
dedicated to the possibility of generating a sizeable baryon asymmetry
on the electro-weak phase transition (see for example 
\cite{nel1,nel2,aya1,far1}). 
For an excess of baryons to develop in an Universe which initially has zero
baryon number, the already well known 
following conditions, first enucianted by Sakharov, must be
met: 
1) Some interaction of elementary particles must violate baryon number. 
2) C and CP must be violated in order that there is not a perfect equality
between
rates of $\Delta B\not= 0$ processes, since otherwise no asymmetry could evolve
from an initially symmetric state. 
3) A departure from thermal equilibrium must
play an essential role, since otherwise CPT would assure compensation between
processes increasing and decreasing the baryon number. 
Remarkably, the standard model of weak interactions may provide all the
necessary
ingredients for baryogenesis. In particular the third condition can be met
if the weak phase transition is at least  weakly first order. 

In a first order  phase transition, the conversion from one phase to
the other occurs trough nucleation. This happens when the system is either 
supercooled or superheated. Bubbles of the ''true''
 phase (with an expectation value
of some Higgs field $v\not =0$) expand rapidly absorbing the region of the
''false'' phase ($v\not=0$). 
At the bubble surface there is a region or ''wall'', in principle of
 microscopic dimensions, which separates the phases. The speed
of the expanding bubble walls could be in the range
 $0.1-0.9 \ c$ \cite{din1}. Particles in the
''false'' (higher temperature) phase are reflected off the advancing bubble walls, while most 
particles in the low temperature phase are unable to catch up with the receding
walls, and cannot equilibrate across the phase boundary. Thus one has a 
departure from equilibrium and a baryon asymmetry can be generated.

In the physical conditions of the early Universe the fermions 
moving through
the bubble wall will interact also 
with the particles in
the surrounding plasma, thus a full transport problem must 
be considered.
A useful simplifying assumption is to decompose the process into 
two steps,
one describing the production of the CP asymmetry 
on the transmission/reflection coefficients
when the 
quarks/antiquarks
are scattered on the wall, the second describing the transport and the eventual
transformation of the CP asymmetry into a baryon asymmetry via the baryon
number anomaly.

Assuming  that the scattering from the wall is little
 affected by diffusion
corrections, the effects of the surrounding plasma can be
 partially incorporated
by introducing a Higgs field effective potential taking into account
finite temperature corrections to the tree-level potential.
The structure of the wall depends on this effective potential on 
a complicated way. 
Fermions passing through the domain wall acquire a mass which is proportional 
to the  vacuum expectation value (VEV) of the Higgs field,
which is  determined from the equations of motion of the
finite temperature effective action of the bubble.
The problem of computation of transmission coefficients reduces to the 
solution of a Dirac equation with a space dependent mass term. 
Exact solutions has been obtained only 
for two simple cases:
for a wall profile approximated by a step function  (\cite{far1,gav1}) and for 
an average Higgs field profile of the type (\cite{nel1}):
\begin{equation}
\phi(z)=\frac{v}{2}\left ( 1+\tanh \left (\frac{z}{\delta}\right )\right )
\end{equation}
where the width of the wall is given by
$$\delta=\frac{2\surd2}{v\surd\lambda}=\frac{\surd 2}{M_H}$$
and $M_H=\sqrt{\lambda}v/2$ is the Higgs mass.
This profile is the analytic solution of 
the Higgs field equation for a bubble
with its normal  along the z-axis and position at $z=0$ under some
simplifying assumptions. 
There is not variation of the complex phase of 
the Higgs field through the wall.
Numerical solutions have been obtained for some
more complicated profiles which incorporate 
locally dependent complex phases and which are 
considered to be ''reasonable'' enough although not neccesarily solution of any
equation of motion ( for example in \cite{nel2}).

For a given  wall profile, 
parallel to the x-y plane and normal to the z-axis
in its own  rest frame,
in order to compute the reflection coefficient one need
only the plane wave solutions of the Dirac equation for particles moving along
the z-axis. For any other incoming direction, the problem can be reduced to the
latter performing an appropriate Lorentz boost.
Following \cite{nel1,rod1}, it is advantageous to work in the chiral basis,
reordering  the spinorial components
the Dirac operator can be factorized into $2\times2$ blocks. For solutions with
positive energy E
\begin{equation}
\Psi=e^{-iEt}\pmatrix{ \psi_I(z) \cr \psi_{II}(z)\cr}\equiv
e^{-iEt}\pmatrix{\psi_1\cr\psi_3 \cr \psi_4 \cr \psi_2}
\end{equation}
where $\psi_1$ and $\psi_2$ are eigenspinors of the chirality operator
, $\gamma_5$, for the eigenvalue $+1$ and $\psi_3,\psi_4$ for $-1$, we
obtain the two equations
\begin{eqnarray}  
(i\partial_z+ Q(z))\psi_I&=&0 \label{e1102}\\
(i\partial_z+ Q^\ast(z))\psi_{II}&=&0 
\label{e1101}
\end{eqnarray}
With 
\begin{equation}
Q(z)=\pmatrix{E & -m(z) \cr m(z)^\ast & -E}
\end{equation}

We will deal in this work with the particular case given by the function
\begin{equation}
m(z)=\left\{ 
\begin{array}{ll}
m_0 \exp i( -\Delta \theta \lambda z+\theta_0) \exp -\lambda z&
 \mbox{if $z>0$} \\
m_0 \exp{i\theta_0} & \mbox{if $z<0$}
\end{array}\right .
\label{e1103}
\end{equation}
All the constants are supposed real. $\lambda >0 $. This function
represents a
linear phase variation with a global
 difference of $\Delta \theta$ over a distance of 
the order of the wall thickness $\delta\equiv 1/\lambda$.

\section{Solving the Dirac equation.}
It would be possible to solve Eqs.(\ref{e1102}-\ref{e1101}) with the function
given by Eq.(\ref{e1103}), reducing them to a  
Bessel-like differential equation. It is possible and advantageous 
however to use perturbation theory to compute the evolution operator of the
system. The summation to all orders of the perturbation expansion is possible
thanks to the special form of $Q(z)$. 
There are two main advantages in doing so:
the first one is that
the procedure is easily generalizable to any number of dimensions (for 
example to incorporate mixing between generations), the second one 
is that the
evolution operator is computed directly and the reflection coefficients
are easily given in terms of its components. 
A similar technique has been used  already in (\cite{emi1,emi2}) to compute 
the neutrino 
oscillation probabilities in solar matter.

The evolution operator of the differential system  (\ref{e1102})
is given by the path-ordered integral
\begin{equation}
U(z,z_0)=P\exp -i\int_{z_0}^z dz Q(z)
\label{e102}
\end{equation}
In this work we are concerned with an operator $Q$ of the form
\begin{equation}
Q=Q^0+  V(z)
\label{e6009}
\end{equation}
With
\begin{equation}
\begin{array}{cc}
Q^0=\pmatrix{E & 0 \cr 0 & -E}; & V=\pmatrix{0 & -k \exp - \sigma z\cr
k^\ast\exp - \sigma^\ast z& 0}
\end{array}
\label{e1110}
\end{equation}

$k,\sigma$ 
are in general complex. The real part of $\sigma$ is greater than zero. It
can be set $\Re \sigma=1$ without loss of generality.

 Formally, it is 
possible to solve  Eq.(\ref{e102}) by successive iterations:

\begin{equation}
U(z,z_0)=U^{(0)}(z,z_0)+\sum_{n=1}^{\infty}U^{(n)}(z,z_0) ;\quad
U^{(0)}(z,z_0)=\exp\left (-i Q^{0}(z-z_0)\right )
\end{equation}

 $U^{(n)}$ is the well-known integral
\begin{equation}
U^{(n)}=(-i)^n\int_{\Gamma}
dz_n
\dots dz_1 
U^0(z,z_n)V(z_n)
\dots
U^0(z_2,z_1)V(z_1)U^0(z_1,z_0)
\label{e101}
\end{equation}

The domain of integration is defined by 
$$\Gamma\equiv z>z_n>\dots >z_1>z_0.$$

Following the same arguments as in \cite{emi1} one can show that it is
enough to compute
U in the $z\rightarrow\infty$ limit.
The evolution for finite time can be deduced from the expression for this
limit. 
 Trought elementary manipulations of Eq.(\ref{e101}), we get the 
elements of $U^{(n)}$ in a basis of eigenvectors of $Q^0$:
\begin{eqnarray}
\lefteqn{<b\mid U^{(n)}\mid a> =  
(-i)^n\exp(-i (Q_b^0 z-Q_a^0 z_0)) \times}\nonumber \\
 & &\times \sum_{k1,\dots,k(n-1)}\int_{\Gamma}
d^n\tau
\exp(iz_n w_{bk1}+\dots+ iz_1 w_{k(n-1)a} ) 
V_{bk1}(z_n)\dots V_{k(n-1)a}(z_1) 
\label{e1107}
\end{eqnarray}
With $w_{k1k2}=Q_{k1}^0-Q_{k2}^0$. $Q^0_k$ one of the two eigenvalues of
$Q^0$.

Due to the  dimensionality of the problem and the special form for
V, the summatory in Eq.(\ref{e1107}) either is zero or reduces to only one term
depending on whether n is odd or even and on the states $a,b$. 
For diagonal terms ($a=b$) the 
 product of $V\dots V$ is always 
zero when n is odd. When n is even there is a single  surviving term.
For non-diagonal terms, the situation is reversed: the
 only one surviving term appears for n odd.
This single 
remaining term is always of the alternating form  
$\dots V_{12}V_{21}V_{21}\dots $. So,
\begin{equation}
<b\mid U^{(n)}\mid a> =  
(-i)^n\exp(-i (Q_b^0 z-Q_a^0 z_0))I_{ab}^{(n)}\times  
\left \{\begin{array}{cc}
\left (-\mid k\mid\right )^{n/2}  &  a=b \\
V_{ba} \left (-\mid k\mid\right )^{(n-1)/2} &  a\not=b 
\end{array}
\right .
\label{e6020}
\end{equation}

All the functions appearing in the integral $I_{ab}^{(n)}$
are of exponential type,
the following equality (\cite{emi1}) can be applied:
\begin{eqnarray}
I_n( w_1,\dots,w_n) &
\equiv &
\int_{z_0}^\infty\dots\int_{z_0}^{x_2} dx_n \dots dx_1\exp
\sum_n w_n x_n \nonumber\\
& =&
   \frac{ (-1)^n \exp (z_0\sum_n w_n)}{w_1 (w_1+w_2)\dots 
(w_1+w_2\dots +w_n)}\nonumber \\
& & \nonumber\\
& & (\hbox{valid if\ }\ \  \Re{\ w_n} < 0,\ \forall\ n\ ) \label{e612}  
\end{eqnarray}

In our case 

\begin{equation}
\begin{array}{rl}
w_1+w_2+\dots + w_j &= i w_{bk1}-L+ i w_{k1k2}-L^\ast+\dots +i
w_{k(j-1)kj}-L\\
 &= iw_{bkj}+n_1 L+n_2 L^\ast
\end{array}
\end{equation}

$w_{bk(j)}$ can take only the values $\{0,\pm 2E\}$. $L$ is
$\sigma$ or $\sigma^\ast$. One have for the integers $n_1,n_2$: 
$n_1=n_2$ or $n_1=n_2\pm 1$.

For the diagonal terms, $a=b$, $n$ even  (writing $s=2Ei-\sigma$): 
\begin{eqnarray}
I_{aa}^{(n),even}&=& 
\frac{\exp z_0 (n/2)(s+s^\ast) }
{s^\ast  (s^\ast+s) (s^\ast+s+s^\ast)\dots((n/2) (s+s^\ast))}
\nonumber \\
&=& \frac{\exp z_0 (n/2)(s+s^\ast) }
{\prod_{j=2,even}^n (j/2) (s+s^\ast) \prod_{j=1,odd}^{n-1} [s^\ast+(j-1)(s+s^\ast)/2]}
\nonumber \\
&=&\frac{\exp z_0 (n/2)(s+s^\ast) }
{(s+s^\ast)^{n/2} \left(\frac{n}{2}\right )! (s+s^\ast)^{n/2}\prod_{l=1}^{n/2} [
s^\ast/(s+s^\ast)+ (l-1)] }
\nonumber \\
&=&\frac{\exp z_0 (n/2)(s+s^\ast) }
{(s+s^\ast)^n \left(\frac{n}{2}\right )!  \left [s^\ast/(s+s^\ast)\right 
]_{(n/2)}}
\label{e6019}
\end{eqnarray}

For non-diagonal terms, taking  $a=1,b=2$ to simplify the notation:
\begin{eqnarray}
I_{ab}^{(n),odd}&=& 
\frac{(-1)\exp z_0 ((n/2)(s+s^\ast)+s^\ast) }
{s^\ast  (s^\ast+s) (s^\ast+s+s^\ast)\dots((n/2) (s+s^\ast)+s^\ast)}
\nonumber \\
&=& \frac{(-1)\exp z_0 (n/2)(s+s^\ast) }
{\prod_{j=2,even}^{n-1} (j/2) (s+s^\ast) \prod_{j=1,odd}^{n} 
(s^\ast+(j-1)(s+s^\ast)/2)}
\nonumber \\
&=&\frac{(-1)\exp z_0 (n/2)(s+s^\ast) }
{(s+s^\ast)^{(n-1)/2} \left(\frac{n-1}{2}\right )! (s+s^\ast)^{(n+1)/2}\prod_{l=1}^{(n+1)/2} [s^\ast/(s+s^\ast)+ (l-1)] }
\nonumber \\
&=&\frac{(-1)\exp z_0 (n/2)(s+s^\ast) }
{(s+s^\ast)^n \left(\frac{n-1}{2}\right )!  \left [s^\ast/(s+s^\ast)\right 
]_{(n+1)/2}}
\label{e6018}
\end{eqnarray}

We have used the Pochammer symbol defined by Eq.~(\ref{a110}).

So, inserting Eqs.(\ref{e6019},\ref{e6018}) in Eq.(\ref{e6020}) and taking $s+s^\ast=-2$, $z_0=0$:

\begin{eqnarray}
\lefteqn{<1\mid U\mid 1>=} \nonumber \\
 &= & 
e^{-i E z}  \left ( 1+\sum_{n=2,even}^{\infty} 
(-i)^n \left (-\mid k\mid^2\right )^{n/2}   
\frac{1}
{(-2)^n (\frac{n}{2})!  [ -s^\ast/2]_{(n/2)} }\right )
 \nonumber \\
 &= & e^{-i E z}
\left ( 1+ \sum_{m=1}^\infty
\left (\frac{\mid k\mid^2}{4}\right )^m \frac{1}{m! [-s^\ast/2]_{(m)}}\right )
 \nonumber \\
&=& e^{ -i E z} 
{}_0 F_1 \left (- \frac{s^\ast}{2}; \frac{\mid k\mid^2}{4}\right )
\end{eqnarray}

\begin{eqnarray}
\lefteqn{<1\mid U\mid 2>=} \nonumber \\
 &= & e^{- i E z}  
(-k)\sum_{n=1,odd}^{\infty} 
(-i)^n \left (-\mid k\mid^2\right )^{(n-1)/2}   
\frac{(-1)}{
(-2)^n (\frac{n-1}{2})! [ -s^\ast/2]_{((n+1)/2)}}
\nonumber \\
&=& e^{-i E z}\frac{2ik}{\mid k\mid^2}   
\sum_{m=1}^{\infty}    
\left (\frac{\mid k\mid^2}{4}\right )^m  
 \frac{1}{\left(m-1\right
)! [-s^\ast/2]_{(m)}} \nonumber \\
&=&e^{- i E z} \frac{ik}
{s^\ast}
{}_0 F_1\left (1-\frac{s^\ast}{2};
\frac{\mid k\mid^2}{4}\right )
\end{eqnarray}
and similarly for the other two matrix elements.
The matrix U can be written as 
\begin{eqnarray}
U(z\to\infty,z_0)&=&\exp-i H_0 (z-z_0) U_{red}( z_0)\nonumber \\
     & & \nonumber \\   
U_{red}(z_0)&=&\pmatrix{
 F &  G \cr G^\ast & F^\ast \cr }
\label{e150}
\end{eqnarray}

with $\sigma=\lambda (1+i\Delta\theta)$, $k=-m_0\exp i\theta_0$,
\begin{eqnarray}
 G&=&\frac{im_0 \exp i\theta_0}{1+(2E-\Delta \theta) i} {}_0 F_1\left (\frac{3}{2}+(E-\frac{\Delta \theta}{2})
 i ; \frac{\mid m_0\mid^2}{4}\right) \nonumber \\
F&=&{}_0 F_1\left (\frac{1}{2}+(E-\frac{\Delta \theta}{2}) i; 
\frac{\mid m_0\mid^2}{4}\right  )
\label{e6024}
\end{eqnarray}
In the case $z_0\not= 0$, $F,G$ would include a factor $\exp -2z_0$ in its
argument. In these formulas $E,m_0$ are given in units of $\lambda$, the 
inverse of
the wall thickness.

To obtain $\overline{U}$, evolution operator 
for the Eq.(\ref{e1101}),  one must make the changes $k\to k^\ast,\Delta\theta\to
-\Delta\theta$. The matrix become
\begin{eqnarray}
\overline{U}(z\to\infty,z_0)&=&\exp-i H_0 (z-z_0) \overline{U}_{red}( z_0)\nonumber \\
     & & \nonumber \\   
\overline{U}_{red}(z_0)&=&\pmatrix{
 \overline{F} &  \overline{G} \cr \overline{G}^\ast & \overline{F}^\ast \cr }
\label{e1150}
\label{e6023}
\end{eqnarray}
with 
\begin{eqnarray}
 \overline{G}&=&\frac{im_0 \exp -i\Delta \theta_0}{1+(2E+\Delta \theta) i} {}_0 F_1\left 
(\frac{3}{2}+(E+\frac{\Delta \theta}{2}) i ; \frac{\mid k\mid^2}{4}\right)\nonumber \\
\overline{F}&=&{}_0 F_1\left (\frac{1}{2}+(E+\frac{\Delta \theta}{2}) i; 
\frac{\mid k\mid^2}{4} \right )
\label{e6025}
\end{eqnarray}

See Appendix A for some new formulas for the absolute values of generalized
hypergeometric functions which can be obtained from the general properties of $U,\overline{U}$.

Following the same reasoning used in (\cite{emi1}), using the
general properties of the evolution operator, the evolution for
any finite z is given by the matrix 
\begin{eqnarray}
 U(z,z_0)&\equiv&U_s(z)^{-1}U_{s}(z_0)=U_{red}^{-1} (z) e^{-i Q^0 (z-z_0)}
U_{red}(z_0) 
\label{e1125}
\end{eqnarray}
In this work, we will make use only of the infinite time limit.

\section{ The reflection Coefficient. Results.}

We will follow the same 
procedure as in (\cite{nel1}) for defining the reflection coefficient.
In the region $z\to\infty$, the eigenspinors $\psi_2,\psi_3$
correspond to right-moving particles with chirality $\gamma_5=+1,-1$
respectively. The solutions $\psi_1,\psi_4$ are identified with
left-moving particles with chirality $+1,-1$.
These states are also eigenstates of the hamiltonian
in this region 

The momentum eigenstates for $z<0$ are obtained diagonalizing
the constant matrix $Q(0)$ by 
\begin{equation}
u(p)=\pmatrix{\cosh\theta_p & \sinh \theta_p\cr \sinh \theta_p& \cosh\theta_p\cr}, \quad
\sinh2\theta_p\equiv \frac{\mid m\mid}{p}=\frac{1}{\sqrt{(E/\mid m\mid)^2-1}}  
\end{equation}
with eigenvalues $p=\pm\sqrt{E^2-\mid m\mid^2}$.

For left-moving particles incident from the
symmetric phase, two components coexist at $z=\infty$, the
incident particle itself and the reflected one by the domain
wall. We define reflection coefficients $R,\overline{R}$ as 
\begin{equation}
\psi_3(\infty)=R \psi_1 (\infty); \quad \psi_2(\infty)=\overline{R}
\psi_4(\infty) 
\end{equation}
Imposing the boundary condition that  at $z<0$ only a left-moving particle with
 momentum $p$ propagates, one gets the expression:
\begin{equation}
\psi(\infty)=U(z\to\infty,0) u^{-1}(p) \psi_p(0)
\end{equation}
where $\psi_p$ is a momentum eigenspinor.
Then, the reflection coefficient is given  by 
\begin{equation}
R=\frac{\left (U u^{-1}\right )_{21}}{\left (U u^{-1}\right )_{11}}; \quad 
\overline{R}=\frac{\left (\overline{U} u^{-1}\right 
)_{21}}{\left (\overline{U} u^{-1}\right )_{11}}
\end{equation}
or more explicitly:
\begin{equation}
R=\frac{U_{21} -t U_{22}}{U_{11}-t U_{12} }=e^{2E zi}\frac{G^\ast-t F^\ast}{F-t G}; 
\quad t=\tanh\theta_p
\end{equation}
and similarly for $\overline{R}$.

\begin{figure}[p]
\centering\hspace{0.8cm}
\epsfig{file=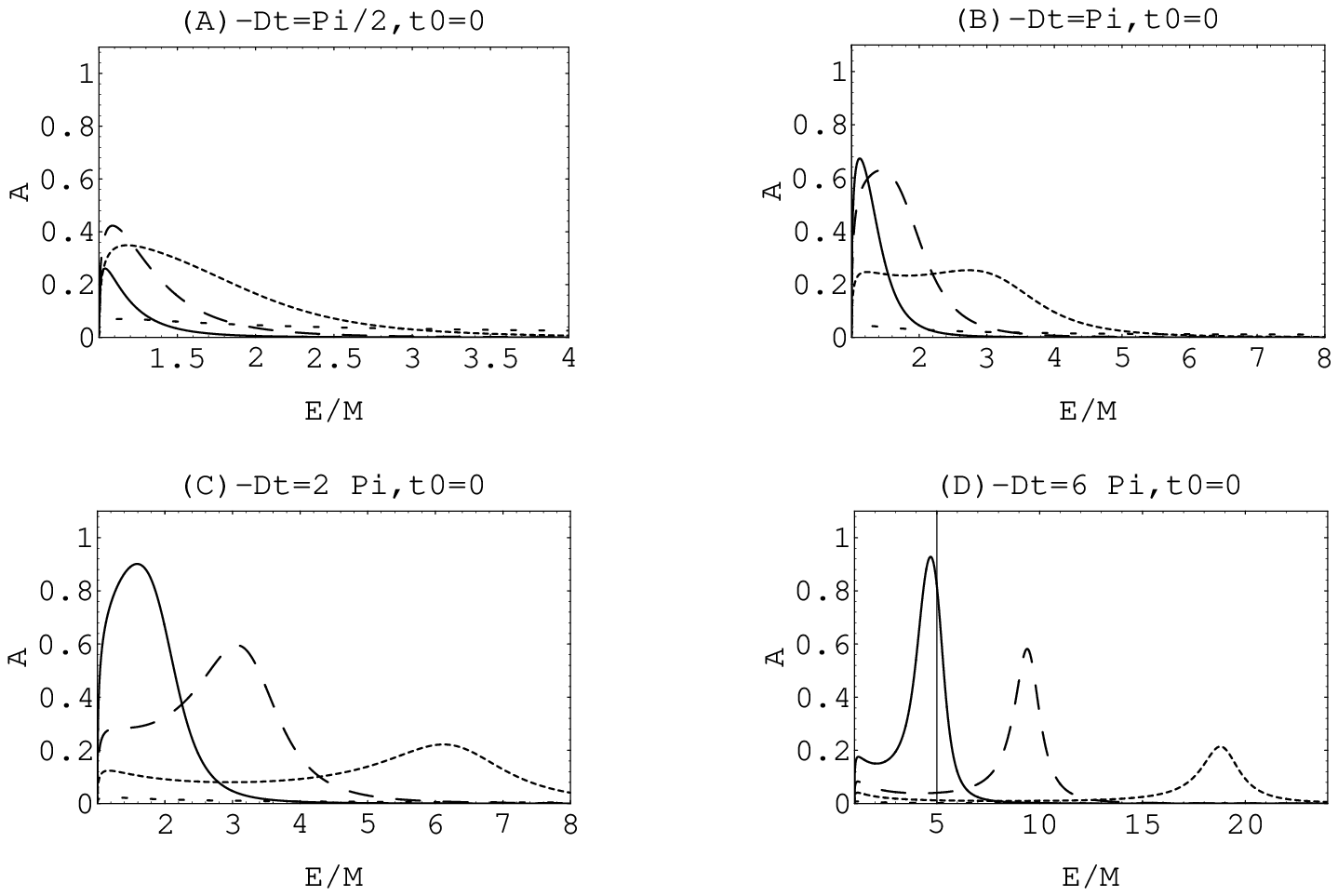,height=14cm}
\caption{The asymmetry A as a function of $E/\mid m\mid$. 
Continuos line: $\mid m\mid \delta=2$, 
dashed lines: respectively 
$\mid m\mid \delta=1,1/2,1/10$. 
From A to C: $\Delta\theta=\pi/2, \pi, 
2\pi, 6\pi$. For all figures $\theta_0=0$.
 }
\label{f1}
\end{figure}

\begin{figure}[p]
\centering\hspace{0.8cm}
\epsfig{file=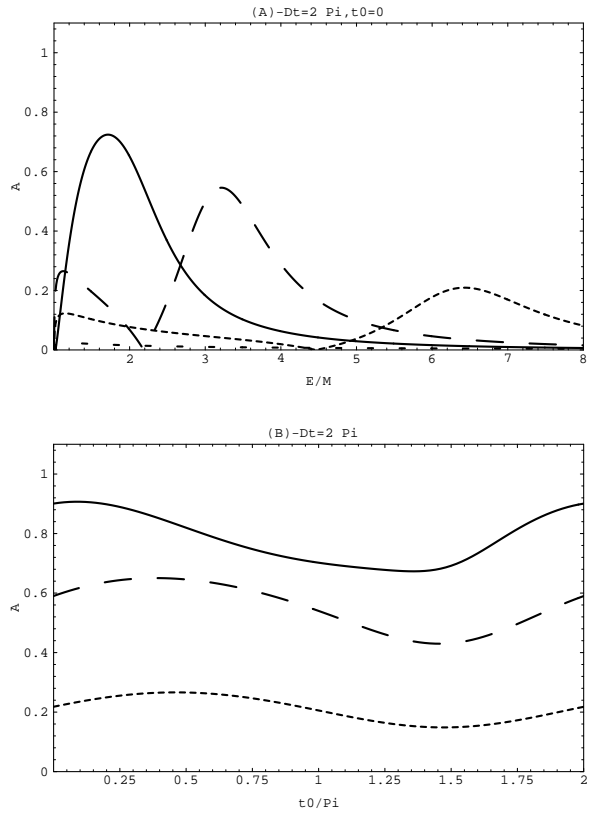,height=12cm}
\caption{Dependence with the initial angle $\theta_0$. Top figure: as in 
Fig.(\protect\ref{f1}) (C) but here for 
$\theta_0=\pi$.
Bottom figure: the quantity $A$
 as a function of $\theta_0$.
$\Delta\theta=2\pi$, $E/ m=\Delta \theta/(2  m\delta)$
(peaks in Fig.(\protect\ref{f1})(C)).}
\label{f2}
\end{figure}

\begin{figure}[p]
\centering\hspace{0.8cm}
{\epsfig{file=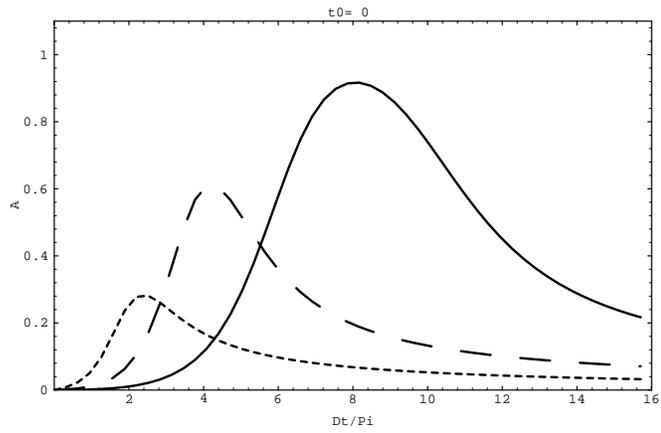,height=12cm}} 
\caption{Asymmetry $A$ as a function of $\Delta\theta$ and 
different $\mid m\mid\delta$ as before. $\theta_0=0$;
$E\delta=2m\delta$.}  
\label{f3}
\end{figure}

We are interested in the asymmetry between the reflection 
probabilities. The quantity $A=\mid R\mid^2-\mid
\overline{R}\mid^2$ 
is displayed in figs.(\ref{f1}-\ref{f3}) as
 a function of the diverse parameters involved $E,\Delta \theta,\theta_0$
and for different values of the product particle mass x wall width $\mid m\mid \delta$.

In Fig.(\ref{f1}) the asymmetry A is plotted as a function of
the dimensionless quantity $E/\mid m\mid$ and some fixed values 
of $\Delta\theta$.
For $\Delta\theta=\pi$ (plot B) 
 the results obtained here coincide with the numerical result obtained in
 \cite{nel1} (Fig. 2).
As the  phase difference $\Delta \theta$ increases, 
the height of the peaks keeps unaltered but their
position moves according to the rule
 $$ (E/m)_{peak}=\mid \Delta\theta\mid/(2m\delta).$$
The first parameter of the hypergeometric
functions in Eqs.(\ref{e6023}-\ref{e6025}) is real for these peak values.
As the value of $\Delta\theta$ gets higher some resonant
 effect becomes apparent, the peaks get sharper and the
asymmetry is only non-negligible around the peaks. This
is particularly evident in the last plot (C). 
This effect could lead to an essentially monochromatic asymmetry
in scenarios with highly oscillatory or random phase differences.

In  Fig.(\ref{f2}) 
we check the influence of the value of the phase $\theta_0$.
The plot (A)  corresponds to plot (C) in Fig.(\ref{f1}). Here we have set
$\theta_0=\pi$. Some new dips appear at lower $E/m$, otherwise the position
and width of the peaks remain unaltered.
In the plot (B) we present the variation of A (peak values at $\theta_0$) in 
a full cycle of $\theta_0$. The phase  $\theta_0$ appears only in
the multiplicative  exponential coefficient of $G$ in Formula (\ref{e6024}). From this, the
soft, quasi-sinusoidal behavior observed in the graphic is understood.

Finally in Fig.(\ref{f3}) 
we plot A as a function of the  parameter $\Delta\theta$, for
$\theta_0$
fixed and for different  values of the product $m\delta$ as before.
As expected the peaks in the asymmetry function appear for values such that
$$\Delta\theta_{peak}=2 E\delta. $$

As conclusion,  we were able to solve the Dirac equation  with a space
dependent complex mass term which, although not a solution
of the equations of motion,  reproduces  
the expected variation in module and in complex phase  of the average Higgs
field across the wall.
From the solution of the Dirac equation the particle-antiparticle 
transmission asymmetry  is computed.
The analytical results presented here confirm previous numerical 
computations and predict an unexpected behavior for highly oscillatory phase
fields.

We note that with little modifications our method can be used  to solve
the Dirac equation for a purely local dependent phase mass term ($\lambda\to 0,
\Delta\theta\lambda\not=0$ in Eq.(\ref{e1103}) or equivalently $\sigma$ purely 
imaginary in Eq.(\ref{e1110})). 
In order to circumvent convergence problems this must be done
  taking the
limit $\Re \lambda\to 0$ in the finite propagation time Eq.(\ref{e1125}).

  \appendix

  \section{Appendix: some old and new formulas about Hypergeometric Functions}

  The generalized Hypergeometric function (\cite{grad}) is defined by
  
\begin{equation}
  {}_p F_q (a_1,\dots,a_p,b_1,\dots,b_q; z)=\sum_{n=0}^\infty 
  \frac{(a_1)_{(n)}\dots (a_p)_{(n)}}{(b_1)_{(n)}\dots (b_q)_{(n)} }
    \frac{z^n}{n!}
  \label{a140}
  \end{equation}

where    the Pochammer symbol is 
\begin{equation}
  (z)_{(n)}=\Gamma(n+z)/\Gamma(z)
  \label{a110}
  \end{equation}

in particular
\begin{equation}
  \mbox{${{}_0 F_1}$}(b;z)=\sum_{n=0}^\infty \frac{1}{(b)_{(n)}} \frac{z^n}{n!}
  \label{a100}
  \end{equation}

The derivative of ${}_0F_1$ is again a ${}_0 F_1$ function:
\begin{eqnarray}
    \frac{d\ \mbox{${{}_0 F_1}$}(\gamma;z)}{dz}&=&\frac{1}{\gamma}\
  \mbox{${{}_0 F_1}$}(1+\gamma;z)  \label{a120} 
\label{a160}
\end{eqnarray}

The function ${}_0 F_1$ is related to the Bessel Functions
by the formula
\begin{equation}
J_n(z)=\frac{(z/2)^n}{\Gamma(1+n)} {}_0F_1\left (1+n; -\frac{z^2}{4}\right )
\end{equation}

For a $Q$ as 
given by Eq.(\ref{e6009}) is traceless, $det \ Q(z,z_0)=det\ Q(0,0)=1$, or
$\mid F\mid^2-\mid G\mid^2=1 $.
The formulas (\ref{e6024}) are valid for any
$E,\Delta\theta$ real, $k$ complex, we obtain:
\begin{equation}
\mid {}_0 F_1\left (\frac{1}{2}+\epsilon i;  x^2\right )\mid^2-\frac{x^2}{\mid
1/2+\epsilon i\mid^2} \mid {}_0 F_1\left (\frac{3}{2}+\epsilon i; x^2\right )\mid^2=1
\end{equation}
to be compared with the similar formulas obtained in (\cite{emi1}) for the
absolute values of the generalized hypergeometric functions ${}_n F_n$.

In fact, with little changes, one could compute U for any general matrix V
with diagonal terms equal to zero and non-diagonal terms of general exponential
type not neccesarily equal.
In the particular case of V hermitic: $V_{12}=V_{21}^\ast=k$,  $U_{red}$
 is unitary and of the form
\begin{eqnarray}
U_{red}(z_0)&=&\pmatrix{
 \overline{F} & \overline{G} \cr -\overline{G}^\ast & \overline{F}^\ast \cr }
\label{e6027}
\end{eqnarray}
with $\overline{F},\overline{G}$ given by Eq.(\ref{e6023}).

By the unitarity of the matrix (\ref{e6027}), 
$\mid F\mid^2+\mid G\mid^2=1 $. And we get the formula:
\begin{equation}
\mid J_{-\frac{1}{2}+\epsilon i}(x)\mid^2+\mid J_{\frac{1}{2}+\epsilon
i}(x)\mid^2=\frac{2\cosh \pi \epsilon}{\pi x}
\end{equation}
For $\epsilon=0$ this formula reduces to the special case
involving the well known 
Bessel functions of order $1/2$: $J_{-1/2}(x)=\sqrt{2/(\pi x)}\cos(x); \
J_{1/2}(x)=\sqrt{2/(\pi x)} \sin(x) $.

\vspace{1cm}
{\Large{\bf Acknowledgments.}}

I would like to thank to Peter Minkowski for many enlightening discussions. This 
work has been supported in part by the Wolffman-Nageli Foundation (Switzerland)
and by the MEC-CYCIT (Spain).

\newpage

\newpage
%

\end{document}